\documentclass[preprint2]{aastex}

\newcommand{\etal}{et al.\ }
\newcommand{\lya}{Ly$\alpha$\ }

\newcommand{\nh}{N_{\rm HI}}
\newcommand{\kms}{\;{\rm km}\,{\rm s}^{-1}}

\newcommand\junits{{\rm erg\,s}^{-1}\,{\rm cm}^{-2}\,{\rm sr}^{-1}\,
		   {\rm Hz}^{-1}}
\newcommand\cdunits{{\rm cm}^{-2}}
\newcommand\ggh{{\Gamma_{\rm HI} }}

\slugcomment{submitted to ApJ}

\shortauthors{Dav\'e \& Tripp}
\shorttitle{Low-$z$ \lya Forest}
\singlespace

\received{2000 November 21}
\begin{document}

\title{The Statistical and Physical Properties of the Low Redshift 
Lyman Alpha Forest Observed with {\it HST}/STIS\footnote{Based on observations 
with the NASA/ESA {\it Hubble Space Telescope}, obtained at the Space Telescope 
Science Institute, which is operated by the Association of Universities for 
Research in Astronomy, Inc., under NASA contract NAS 5-26555.}}

\author{Romeel Dav\'e \altaffilmark{2,3} }
\affil{Steward Observatory, University of Arizona, Tucson, AZ 85721}
\and
\author{Todd M. Tripp \altaffilmark{4} }
\affil{Department of Astrophysical Sciences, Princeton University,
    Princeton, NJ 08544}

\altaffiltext{2}{rad@as.arizona.edu}
\altaffiltext{3}{Hubble Fellow}
\altaffiltext{4}{tripp@astro.princeton.edu}

\begin{abstract}

We examine the \lya absorber population at $z <$ 0.3 detected in
spectra of the QSOs PG0953+415 and H1821+643 taken with the Space
Telescope Imaging Spectrograph aboard the {\it Hubble Space Telescope}.
We compare their statistical properties to those in carefully-constructed
mock quasar spectra drawn from a cosmological hydrodynamic simulation of a
$\Lambda$CDM universe.  We find very good agreement in the column density
and $b$-parameter distributions, down to the smallest observable absorbers
with $\nh\approx 10^{12.3}\cdunits$.  The observed absorber population
is complete for $\nh\ga 10^{13}\cdunits$, with a column density
distribution slope of $\beta=2.04\pm 0.23$ and a median $b$-parameter of
21~km/s above this limit.  The intergalactic gas giving rise to these
weak absorbers is analogous to that at high redshift, located in diffuse
large-scale structures that are highly photoionized by the metagalactic
UV flux, though a greater number arise within shock-heated warm gas.
The density, temperature, and column density of these absorbers follow
similar relationships to those at high redshift, though with substantially
larger scatter due to the shock-heated gas.  The $b$-parameters typically
have a significant contribution from thermal broadening, which facilitates
a measurement of the low-$z$ IGM temperature as traced by \lya absorbers.
From our simulation we estimate $T_{\rm IGM}\sim 5000$~K, with an upper
limit of $10^4$~K, at the mean density.  The agreement in predicted
and observed amplitude of the column density distributions allows us
to measure the \ion{H}{1} photoionization rate at $\bar{z}=0.17$ to be
$\ggh\; =10^{-13.3\pm 0.7} {\rm s}^{-1}$ (estimated modeling uncertainty),
close to predictions based on quasar properties.
\end{abstract}


\section{Introduction}

Neutral hydrogen along the line of sight to distant quasars produces
numerous Lyman alpha (Ly$\alpha$) absorption features in quasar spectra,
known as the \lya ``forest" \citep{lyn71}.  At redshifts $z\ga 2.5$,
high-resolution optical spectroscopy with HIRES \citep{vog94} on
the Keck 10m telescope has enabled the statistics of \lya absorbers
to be determined to high precision \citep{hu95,lu96,kim97,kir97}.
Recent observations using UVES at the VLT have probed absorbers
at redshifts $1.5\la z\la 2.5$ \citep{kim00}.
In conjunction with these observations, hydrodynamic simulations
of the intergalactic medium (IGM) have forwarded a self-consistent
physical picture for the origin of \lya forest absorbers ($\nh\la
10^{17}\cdunits$), in which they arise from highly photoionized gas
tracing the dark matter potentials in mildly nonlinear large-scale
structures \citep[for a review, see][]{rau98}.  Though the absorbing gas
is typically out of dynamical and thermal equilibrium, a relatively
simple relation known as the Fluctuating Gunn-Peterson Approximation
\citep[FGPA; for the full expression see][]{cro98} captures the essential 
physics of optically-thin \lya absorbers:
\begin{equation}\label{eqn: fgpa} 
\tau \propto \rho^{1.6}\ggh^{-1}, 
\end{equation} 
where $\tau$ is the \ion{H}{1}
optical depth, $\rho$ is the local density of baryons (assumed to trace
the dark matter), and $\ggh$ is the \ion{H}{1} photoionization rate.
The FGPA has been utilized in various forms to constrain many physical
parameters associated with high-redshift \lya forest absorption, such
as a measurement of $\Omega_b$ \citep{rau97,wei97} and the
mass power spectrum at $z\sim 3$ \citep{cro98,mcd00}, and
implicitly in the determination of the metallicity of the diffuse IGM
\citep{rau97a,dav98} and the ``equation of state" of the IGM
\citep{hg97,sch99,mcd00b}.

At redshifts $z\la 1.5$, the \lya transition falls in the ultraviolet,
requiring more challenging space-based observation.  The {\it Hubble Space
Telescope (HST)} Quasar Absorption Line Key Project \citep{bah93,bah96}
obtained UV quasar spectra for a large sample of quasars \citep{jan98}
using the {\it HST} Faint Object Spectrograph (FOS).  Unlike Keck/HIRES,
FOS is unable to resolve individual \lya absorption features, having
a resolution of $\approx 230$~km/s.  Nevertheless, this sample greatly
expanded our understanding of the properties and evolution of \lya forest
absorbers from $z\sim 1.7\rightarrow 0$.  One surprising result was that
the rate of evolution of absorbers slows dramatically at $z\la 2$; this
was initially indicated by the large number of Ly$\alpha$ lines detected
in the first observations of 3C 373 \citep{mor91,bah91} and confirmed with
a larger number of sight lines from the FOS Key Project \citep{bah96}.
Cosmological hydrodynamic simulations suggested that this was a result of
the diminishing ionizing background in concert with the decline in the
quasar population after $z\sim 2$ \citep[hereafter DHKW]{the98,dav99}.
FOS data also suggested that high-column density absorbers are evolving
away faster than low-column density ones \citep{wey98}, which DHKW
suggested was due to the increase in the relative cross-section
of lower-column absorbing gas.  DHKW also showed that the observed
correlation between absorber equivalent width and impact parameter
\citep{che98,tri98} arises purely due to the clustering of matter,
not due to a physical association of absorbers with \ion{H}{1} in large
gaseous halos of individual galaxies.  Thus qualitatively, \lya absorbers
at low redshift are well-described by current structure formation models.
Still, due to the low resolution of FOS and the computational difficulty
of running simulations to $z=0$, a detailed quantitative agreement of
simulations with observations analogous to that obtained at high redshift
has yet to be achieved.

{\it HST's} Goddard High Resolution Spectrograph (GHRS) has yielded
some complementary results regarding low-$z$ \lya absorbers.  In the
intermediate-resolution mode usually employed in QSO absorption line
studies, the GHRS provided an instrumental resolution of $\sim$19~km/s,
which is comparable to the widths of \lya lines, making a determination
of individual absorber column densities and $b$-parameters possible,
though not straightforward \citep{pen00}.  The {\it Far Ultraviolet
Spectroscopic Explorer (FUSE)} can be used to detect higher Lyman series
lines at wavelengths shortward of the {\it HST} bandpass and thereby
enable a curve-of-growth analysis to more accurately estimate line
widths \citep{shu00}.  The resulting widths were typically half
of that obtained by direct profile fitting of GHRS data, which
\citet{shu00} interpreted as an indication of non-thermal components in \lya
absorbers.  

The deployment of the Space Telescope Imaging Spectrograph (STIS) aboard
{\it HST} has significantly improved our ability to study low-redshift
\lya absorbers.  The intermediate-resolution FUV echelle mode of STIS has
a resolution of 7~km/s FWHM and enables detection of absorbers with $\nh
\leq 10^{13}\cdunits$.  This resolution, similar to that of Keck/HIRES,
allows virtually all \lya absorbers to be fully resolved, facilitating a
robust comparison with high-resolution hydrodynamic simulations.  In this
paper we determine the column density and linewidth distributions from
STIS spectra of quasars PG~0953+415 ($z_{\rm em}$ = 0.239) and H1821+643
($z_{\rm em}$ = 0.297), and we compare them to carefully-constructed
mock quasar spectra drawn from a cosmological hydrodynamic simulation of
a $\Lambda$CDM universe.  We find that the agreement between simulations
and observations is quite good, and lends strong support to the physical
picture of the low-redshift intergalactic medium provided by hierarchical
structure formation models.

DHKW suggested that the physical state of the gas giving rise
to \lya absorption at low redshift is fundamentally similar to that at
high redshift, with the exceptions that a dynamically equivalent patch
of gas at low redshift gives rise to an absorber with significantly
lower column density (see their Figure~10), and collisionally ionized
hot gas plays an increasingly important role to lower redshift.  With
STIS data, we can now probe these low-$\nh$ absorbers that are
predicted to be physically similar to the high-$z$ forest absorbers.
If the FGPA provides a reasonable description of weak low-$z$
absorbers, it may be possible to apply many of the same techniques
used to determine physical properties of the IGM at high redshift to
STIS data.  In this paper we present such applications, including an
estimate of the extragalactic \ion{H}{1} photoionization rate.  Despite
large systematic uncertainties in the modeling, our measurement is
among the most sensitive to date.

In \S\ref{sec: sims} we describe the STIS observations of the two
low-redshift quasars, the hydrodynamic simulations we will use for
comparison, and the construction and analysis of artificial spectra.
In \S\ref{sec: stats} we present the column density and line
width distribution from STIS data alongside the distributions from
carefully-constructed artificial spectra.  In \S\ref{sec: phys} we
explore the physical state of gas giving rise to the sorts of absorbers
seen in these STIS spectra, and make a preliminary determination of the
IGM temperature from the $b-\nh$ distribution.  In \S\ref{sec: ggh} we
constrain the average metagalactic photoionizing flux incident on these
absorbers by matching the amplitude of the column density distribution
to observations.  We present our conclusions in \S\ref{sec: disc}.

\section{Simulation and Observations}\label{sec: sims}

\subsection{STIS Quasar Spectra}

The STIS observations of PG0953+415 and H1821+643 were obtained by
\citet{tri00a} and \citet{tri00b} to study low-redshift
\ion{O}{6} absorbers as well as the relationships between the various
types of absorption systems and galaxies. After the publication of
those papers, more observations were obtained, and modest improvements
in the data reduction procedures were implemented. A detailed
description of the observations, data reduction, and complete line
lists are found in a separate paper \citep{tri01}. Here we provide a
brief summary.

Both QSOs were observed with the medium resolution FUV echelle mode
(E140M).  H1821+643 was observed with the $0\farcs 2 \times 0\farcs 06$
slit, while PG0953+415 was observed with the $0\farcs 2 \times 0\farcs
2$ slit for better throughput. This STIS mode provides a resolution of
$R = \lambda / \Delta \lambda \approx$ 46,000 (7 km/s FWHM) with the
$0\farcs 2 \times 0\farcs 06$ slit \citep{kim98}; the FWHM is only
slightly degraded with the $0\farcs 2 \times 0\farcs 2$ slit, but the
broad wings of the line spread function are more prominent (see Figure
13.87 in the STIS Instrument Handbook, v4.1). This mode provides
spectra extending from 1150 $-$ 1710 \AA\ with only a few small gaps
between orders at $\lambda >$ 1630 \AA , redwards of the \lya forest.

The data were reduced with the STIS Instrument Definition Team
software. The individual spectra were flatfielded, extracted, and
wavelength and flux calibrated with the standard techniques. Then a
correction for scattered light was applied using the method developed
by the STIS Team, and the individual spectra were combined with
weighting based on signal-to-noise ratio (S/N).  Finally, overlapping
regions of adjacent orders were also coadded weighted by S/N.

\subsection{Generation of Artificial Spectra}

We employ a cosmological hydrodynamic simulation of a $\Lambda$-dominated
cold dark matter model, run with PTreeSPH \citep{dav97b}, having
$\Omega_\Lambda=0.6$, $\Omega_{\rm CDM}=0.3527$, $\Omega_b=0.0473$,
$H_0=65 \kms {\rm Mpc}^{-1}$, inflationary spectral index $n=0.95$,
and rms fluctuation amplitude $\sigma_8=0.8$.  We use $64^3$
dark matter particles and $64^3$ gas particles in a periodic cube
11.111$h^{-1}$~comoving Mpc on a side and a gravitational softening
length of 3.5$h^{-1}$~comoving kpc (equivalent Plummer softening).
Further details may be found in DHKW.  We choose $\Omega_b$ to
match recent observations of the deuterium abundance \citep{bur98},
and adopt a UV background from \citet{haa96}.  In post-processing we
adjusted the intensity of the UV background slightly in order to match
the evolution of the number density of absorbers ($dN/dz$) with rest
equivalent width $W_r>0.24$\AA\ (the ``matched LCDM" curve in Figure~4 of
DHKW), by assuming the optical depth in artificial spectra is inversely
proportional to the photoionization rate.  This simulation is otherwise
the same as the ``LCDM" run in DHKW.

From this simulation we generate artificial spectra whose characteristics
closely model those of PG0953+415 and H1821+643; see \citet{dav98}
for a more detailed description of this procedure.  We assume zero
metallicity, so all features in the artificial spectra are Lyman series
lines.  The flux is interpolated to the wavelength of each pixel in the
observed spectrum, and Gaussian random noise is added based on the noise
value at the nearest pixel in the observed spectrum with a similar flux.
We convolve the spectrum with the STIS line spread function of the slit
used to observed each quasar, interpolated to the pixels' wavelength.
We fit a continuum to both observed and artificial spectra using a median
filter in a sliding 100-pixel ($\approx 300$~km/s) window.  We adjust
the median filter so that in each set of artificial spectra, there is on
average no net flux subtracted by the continuum fitting (i.e. we recover
the true continuum, on average).  We generate 20 artificial spectra for
each quasar, for a total of 40.

We fit Voigt profiles to the observed and artificial spectra using
AutoVP \citep{dav97a}.  We choose a $4.5\sigma$ detection threshold for
identifying \lya absorbers, using the procedure described in \citet{lan87}
with a 20~km/s detection window.  We only consider lines between the \lya
and Ly$\beta$ emission peaks so as to reduce confusion with higher-order
Lyman series and metal lines, and do not include any absorbers within
5000~km/s of the \lya emission peak.  Spectral regions containing
Galactic ISM lines or extragalactic metal lines identified in the
data are discarded, and any lines identified in the artificial spectra
in those wavelength regions are discarded as well, ensuring that each
artificial spectrum has the same redshift path length as its corresponding
quasar spectrum.  Note that considerable care must be exercised in the
identification of the Milky Way features since the spectra have sufficient
sensitivity to show weak ISM lines such as the \ion{C}{1} and \ion{Ni}{2}
resonance transitions; similarly, several extragalactic metal lines are
present in the wavelength range of interest \citep[see][]{tri00b,tri01}.

\begin{figure}
\epsscale{1.2}
\plotone{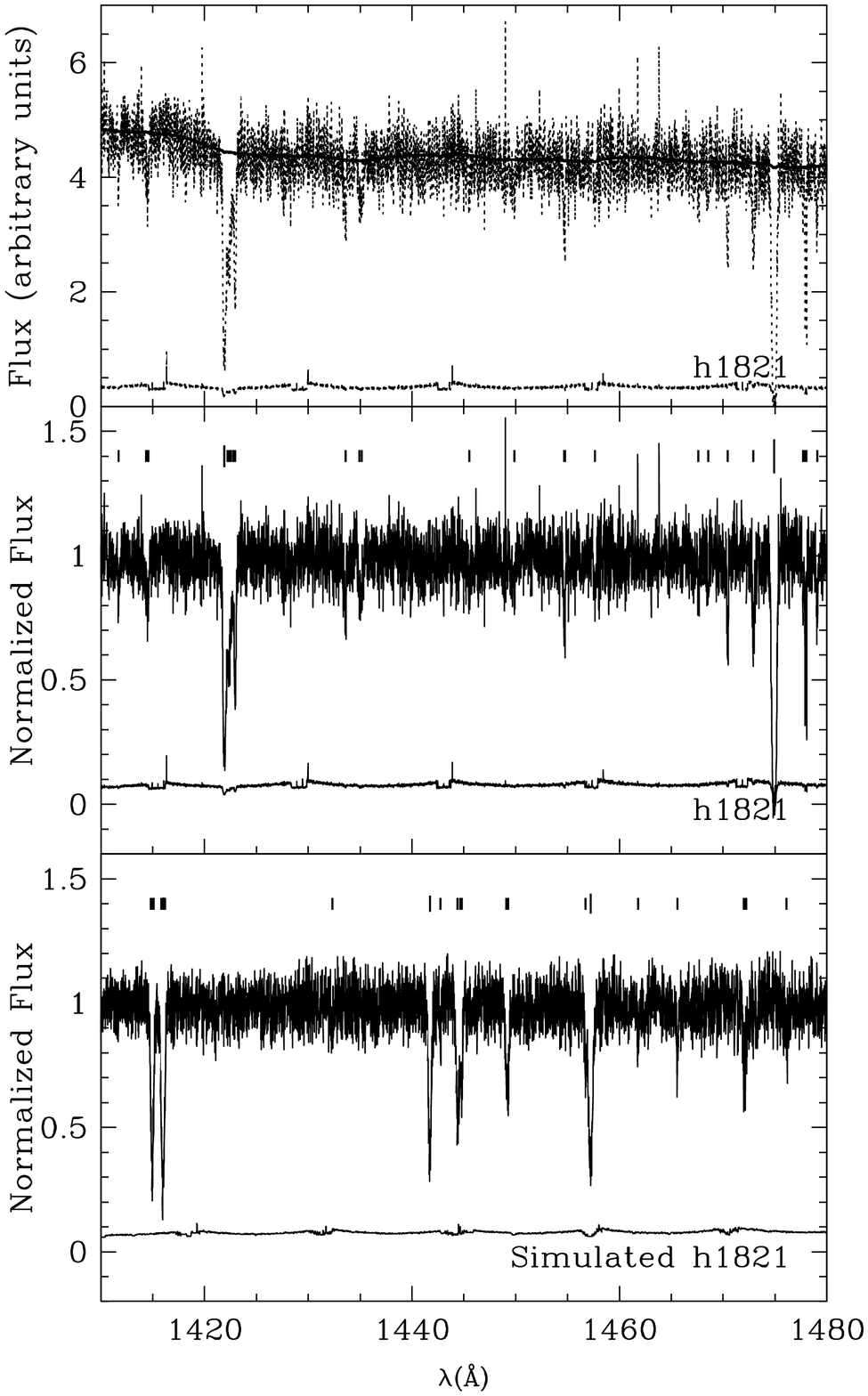}
\caption {An 80\AA\ segment from the \lya forest of H1821+643.
Top panel shows the unnormalized spectrum with continuum fit.   Middle
panel shows the normalized spectrum with line identifications by AutoVP.
Note that some of the marked lines in this panel are not \lya lines, e.g.,
the \ion{Si}{3} 1206.5\AA\ lines at z = 0.225 \citep[see Figure~2 in][]{tri00b} and
the Milky Way \ion{Ni}{2} line at 1454.8\AA.  Bottom panel shows an artificial
H1821+643 spectrum mimicking its resolution and noise characteristics,
continuum fit and analyzed with the same routines applied to the
H1821+643 spectrum.  The $1\sigma$ noise vector is shown near the bottom
of each panel. }
\end{figure}

An 80\AA\ sample of H1821+643 is shown in Figure~1, at three stages.
The first (top panel) shows the unnormalized observed spectrum, with the
continuum fit as the thick solid line, and the $1\sigma$ noise vector at
the bottom.  The middle panel shows the normalized spectrum with Voigt
profile identifications indicated by the tick marks (some of the marked
lines are Milky Way or extragalactic metal lines which are discarded).
The bottom panel shows the corresponding segment from an artificial
H1821+643 spectrum, with the same analysis procedure having been applied.

While there may be biases associated with our particular continuum
fitting, line identification, and Voigt profile fitting algorithms, the
application of the identical routines to both simulations and
observations facilitates a fair comparison between the two.  We do not
expect these biases to be significant since we are fully resolving
virtually all \lya absorbers, and continuum fitting is relatively
straightforward since the low-$z$ forest is sparse.  Any systematic
biases that do arise will, in most cases, be reflected equally in the
statistics of both the simulations and observations.

Our $11.111 h^{-1}$Mpc simulation volume is small compared to the scale
of linearity at $z=0$, hence our derived statistics are subject to
cosmic variance.  We do not attempt to estimate this, since to do so
would require running a suite of simulations that is beyond our current
capabilities.  We note that our randomly chosen volume is typical,
containing no anomalously large mass concentrations or voids.  As we will
show, most \lya absorbers arise in moderate overdensity regions that are
well-sampled even within our small volume, though the contribution from
coupling to waves larger than our boxsize is not included.  We leave
a systematic study of these effects for future work.

\section{Absorber Statistics}\label{sec: stats}

The distributions of column densities and Doppler widths $b$ obtained
by fitting Voigt profiles to spectral features have historically
been used to characterize the statistics of \lya forest absorbers.
Though these statistics reflect the old paradigm that \lya absorbers
arise from isolated thermally-broadened clouds, an assumption that is
likely to be incorrect in detail \citep[e.g.][]{her96,out00,the00},
such statistics still represent a fair statistical characterization of
the absorber population, more so at low redshift than at high redshift
because line blending is less severe and individual absorbers can more
easily be identified.

At high redshift, statistics based on the distribution of flux in the
\lya forest have gained favor \citep[e.g.][]{mcd00,mac00} due to their
simplicity, ease of implementation, and robustness of comparison
with models.  At low redshift, these statistics are somewhat more
difficult to apply reliably, because they require that the opacity in
the forest be predominantly due to \lya absorption, whereas at low-$z$
the relative contribution from metal and ISM lines is greatly increased.
Thus we will leave flux statistics for future work, and focus here
on Voigt profile statistics.

We identify 33 \lya absorption lines in the \lya region of PG0953+415,
and 56 in H1821+643, using AutoVP.  There are more absorbers in H1821+643
because the average signal-to-noise ratio per $\approx 3$~km/s pixel
in this wavelength range is higher, S/N$=13.7$ compared to $8.1$ for
PG0953+415, so more weak absorbers are indentified.  The total redshift
path length is $\Delta z=0.165$ for PG0953+415 and $\Delta z=0.170$ for
H1821+643, not including discarded regions that cover $\Delta z=0.008$ and
$\Delta z=0.011$, respectively.  This path length is for absorbers above
our completeness limit, which we will show is $\nh\ga 10^{13}\cdunits$;
for weaker absorbers, the effective path length is smaller \citep[see,
for example, the discussion in][]{pen00}, but by construction is still identical
in the data and the artificial spectra.  We will only quote statistics
for absorbers above our completeness limit.  The mean redshift of our
entire absorber sample is $\bar{z}=0.17$.

\subsection{Column Density Distribution}

The column density distribution is defined as the number of lines per unit
redshift $z$ per unit column density $\nh$, and is generally parameterized
as a power law in $\nh$,
\begin{equation}\label{eqn: cddist}
f(\nh)\equiv {d^2 N\over dzd\nh}\propto \nh^{-\beta}.
\end{equation}
At $z\ga 2.5$, $\beta\approx 1.5$ \citep{hu95,lu96,kir97,kim97}.
Recent work with VLT/UVES suggests a similar slope ($\beta\approx
1.4$) down to $z\approx 1.5$ \citep{kim00}.  This is predicted to
steepen significantly at low redshift \citep[DHKW]{zha95,the98}.
GHRS observations by \citet{pen00} indicate a somewhat steeper slope of
$\beta=1.8$, but that was obtained by assuming a particular $b$ for all
absorbers (varied between 20 and 30~km/s) rather than by determining $b$
and $\nh$ independently from profile fitting.  An alternative statement
of this trend is that high-$\nh$ absorbers are predicted to evolve away
faster than low-$\nh$ ones.  The analysis in \S 5.2 of \citet{pen00}
suggests no significant difference between the evolution of absorbers
with $10^{13.1}<\nh <10^{14}\cdunits$ and $\nh\ga 10^{14}\cdunits$,
contradicting FOS results \citep{wey98} as well as model predictions.
However, Figure~18 of \citet{pen00} suggests that between $10^{13}\la
\nh\la 10^{14}\cdunits$, stronger absorbers evolve away significantly
faster than weaker ones from $z=3\rightarrow 0$.

\begin{figure}
\epsscale{1.2}
\plotone{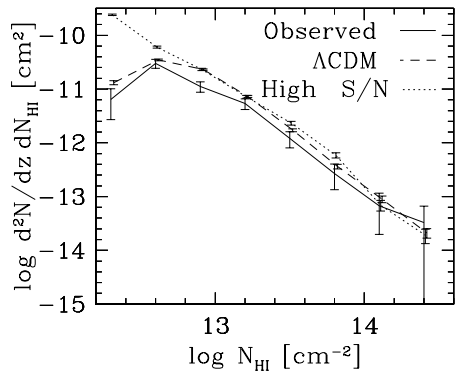}
\caption{Column density distributions from observations of
PG0953+415 and H1821+643 (solid line) and a sample of 20 artificial spectra for
each quasar (dashed line).  Dotted line shows the result from artificial
spectra with S/N increased by $\times 5$, indicating that the absorber
population is complete down to at least $\nh\approx 10^{12.9}\cdunits$. 
Note that the ionizing background used to generate the artificial spectra
was normalized to the evolution of absorbers in the Key Project FOS
sample (see DHKW), and is not the the normalization that best fits this
STIS sample; we will quantify this further in \S\ref{sec: ggh}.  }
\end{figure}

Figure~2 shows the column density distribution of identified absorbers in
the two STIS spectra (combined) along with the corresponding results from
40 artificial spectra.  The simulations and observations show very
good agreement, in both the slope and the amplitude of $f(\nh)$.  The
best-fit slopes for $\nh>10^{13}\cdunits$ are $\beta_{\rm obs}=2.04\pm
0.23$ and $\beta_{\rm sim}=2.15\pm 0.04$.  Thus we find that the the
column density distribution has steepened considerably since $z\sim 3$,
at least for absorbers with $10^{13}\la \nh\la 10^{14}\cdunits$, in general
agreement with Figure~18 of \citet{pen00}.  We do find a slope that is
somewhat steeper (by $\sim 1\sigma$) than that found from the GHRS sample
\citep{pen00}.  Note that we only include $\nh\geq 10^{12.9}\cdunits$ bins
in this fit; had we included the bin at $\nh=10^{12.6}\cdunits$,
the slopes would reduce to $\approx 1.7$.  This bin is only slightly
incomplete (as we discuss below), but it illustrates that completeness
must be carefully assessed before an accurate determination of $\beta$
can be made.

The dotted line in Figure~2 shows the column density distribution for a
set of artificial spectra where the S/N ratio has been increased by a
factor of 5, allowing weaker absorbers to be identified.  No significant
differences are seen for $\nh\geq 10^{12.9}\cdunits$.  This tests
the completeness of the observations (and simulations) at low $\nh$,
showing that the absorber sample in PG0953+415 and H1821+643 is complete
down to $\nh\approx 10^{13}\cdunits$, and $\ga 70$\% complete to $\nh\approx
10^{12.6}\cdunits$.  The similar shape of the turnover below this value
indicates that the noise level in the artificial spectra are faithfully
mimicking that in the STIS data.

The agreement in amplitude of $f(\nh)$ depends on the chosen value of
the \ion{H}{1} photoionization rate used to generate the artificial
spectra, as we will explore further in \S\ref{sec: ggh}.  The best-fit
amplitudes for the simulated and observed absorbers are $f_{\rm
sim}(10^{13}\cdunits)= -10.70\pm 0.02$ and $f_{\rm obs}(10^{13}\cdunits)=
-10.87\pm 0.12$ respectively.  Recall that the ionizing background was
set to obtain agreement with $dN/dz$ of absorbers with $W_r>0.24$\AA\
($\nh\ga 10^{14}\cdunits$) measured by the Key Project's sample of
FOS quasar spectra \citep[see][]{dav99}, which probes a different column
density regime than these STIS spectra.  The general agreement between
these independent data sets lends further support to the overall model.

\subsection{Distribution of $b$-parameters}\label{sec: bpar}

The $b$-parameter distribution has historically provided a significant
challenge for theories of the \lya forest.  At high redshift, absorbers
are predominantly broadened by bulk flow \citep[e.g.][]{her96,out00},
but thermal broadening is not negligible \citep{the00}.  Hence the
$b$-parameter distribution can probe the temperature of \lya forest gas
\citep{sch99,mcd00b,ric00}.  The derived temperatures seem to suggest
that intergalactic gas at $z\sim 3$ is heated by some other process
in addition to equilibrium \ion{H}{1} photoionization, such as helium
reionization \citep{hae98}.  Though some observational constraints are
beginning to emerge \citep{heap00,son96,gir97}, theories for when and
how reionization occurs are quite uncertain, making an {\it a priori}
prediction of $b$-parameters difficult.  Furthermore, it has been shown
that a robust prediction of the $b$-parameter distribution for comparison
with Keck/HIRES data requires exceptionally high numerical resolution
\citep{the98b,bry99}, higher for instance than in the simulation
analyzed here.

At low redshift, it is unclear whether simulation predictions of
$b$-parameters are more or less robust.  Certainly reionization is
far enough in the past that the IGM is expected to have returned to an
equilibrium ionization condition.  However, the character of the low-$z$
IGM has changed dramatically.  In particular, though a substantial
fraction of the baryons remains in cool, photoionized gas, a significant
component of intergalactic gas now resides at temperatures exceeding that
from photoionization, due to shock heating from infall onto large-scale
structure \citep[DHKW;][hereafter D01]{cen99,dav00}.  Furthermore, it is
possible that non-thermal processes such as supernova heating also heat
the IGM \citep[though see D01 and Croft \etal 2000]{wu00}.  The simulation
used here includes these physical processes (it is simulation ``D2" in
D01), but they are more complex than just photoionization heating and
adiabatic cooling, involving highly uncertain theoretical issues such as
escape fraction of feedback energy and the contribution of small-scale
shocks to IGM heating (see D01 for discussion).  Furthermore, the baryonic
mass fraction in various intergalactic components depends on $\Omega_b$
and cosmology (DHKW) in a way that is difficult to predict analytically,
as we will discuss in \S\ref{sec: ggh}.

\begin{figure}
\epsscale{1.2}
\plotone{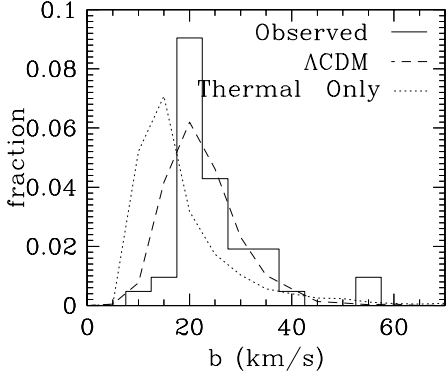}
\caption{$b$-parameter distribution from observations (solid line) 
and artificial spectra (dashed line).  The dotted line shows the distribution
of thermal widths for the absorbers in the artificial spectra, discussed
in \S\ref{sec: temp}.  Only lines with $\nh>10^{13}\cdunits$ are included.  }
\end{figure}

With these caveats in mind, we present in Figure~3 a comparison of
the $b$-parameter distribution in our two STIS quasar spectra and the
artificial spectra.  We only show the distribution for absorbers above
our completeness limit of $\nh>10^{13}\cdunits$, but including absorbers
down to $\nh=10^{12.6}\cdunits$ makes little difference in the results.
Remarkably, despite all the potential pitfalls, the agreement between
simulations and observations is quite good.  The median $b$-parameter
is observed to be 22~km/s, while the simulations predict 21.5~km/s.
The average $b$ is observed to be 25~km/s, and predicted to be 23~km/s. A
K-S test shows no statistically significant difference between the $b$
distributions from the observations and simulations.  These quoted
values include the contribution from the STIS line spread function;
correcting for this lowers these values by $\sim 1$~km/s, yielding an
intrinsic median $b$-parameter of $\approx 21$~km/s.  The dotted line
shows the distribution of thermal widths for the artificial sample,
which we will discuss in \S\ref{sec: temp}.

The median $b$ value is lower than the $27-36$~km/s typically found at
$z\sim 2.5-4$ \citep{kim97}, though a higher S/N spectrum finds somewhat
lower $b$ values \citep{kir97}.  This is qualitatively consistent with the
prediction that the unshocked IGM is cooler at low-$z$ due to the lowered
intensity of the photoionizing background, but could also be a result of
lowered bulk flow broadening due to differential Hubble expansion across
the absorbing structures.  We will explore the relative importance of
these effects in \S\ref{sec: temp}.

Our $b$-parameters do not follow an extrapolation of the trend seen by
\citet{kim97}, that $b$ generally increases with decreasing redshift for
$2.1<z<4$.  They are also significantly smaller than $b$-parameters
derived from profile fitting GHRS data \citep[$\bar{b}=38\pm
16$~km/s,][]{pen00}.  On the other hand, we are in agreement with a
curve-of-growth analysis using FUSE and GHRS data by \citet{shu00}, who
find $\bar{b}=31\pm 7$~km/s and $b_{\rm median}=28$~km/s.  As we will
discuss further in \S\ref{sec: temp}, larger column density absorbers
are wider, and Figure~6 shows that at the median $\nh=10^{14.2}\cdunits$
of the \citet{shu00} sample, our median $b$-parameter is $\approx 30$.
This agrees quite well with their value of 28~km/s, given the small
samples involved.  Our STIS data probes weaker lines than the FUSE sample,
and hence results in a smaller median $b$.

\citet{shu00} find that, for partially saturated lines, direct profile
fitting of GHRS data overstimates the $b$-parameter by a factor of two on
average, and thereby calls into question the validity of profile fitting.
It is true that results from profile-fitting a saturated \lya line can
have large uncertainties.  However, the vast majority of our lines are not
saturated ($\nh\la 10^{14}\cdunits$), so our statistics, particularly
our median $b$ value, should not be too sensitive to this effect.
Furthermore, as evidenced by our agreement with the curve-of-growth
study of \cite{shu00}, we are at most only slightly overestimating
$b$-parameters.  The 7~km/s resolution of our data insures that we
can adequately sample individual \lya absorbers profiles, since the
smallest \lya absorber widths are significantly larger.  Note that we
do find a number of metal and ISM lines that have widths comparable to
the resolution \citep{tri00b,tri01}, indicating that if such \lya lines
existed, they would have been detected.

The agreement of the predicted and observed $b$-parameter distributions
is actually better than that obtained in a similar comparison at high
redshift \citep{dav97a}.  This surprising result may be because STIS
observations do not yet have high enough S/N to show discrepancies with
the simulations.  Or it may be that since low-$z$ forest requires less
deblending of absorption features, the $b$-parameter distribution is more
robustly predicted.  Alternatively, we may have simply been fortunate with
this simulation, and future higher-resolution, larger-volume simulations
will show discrepancies.  For now we are encouraged by the good agreement,
and will proceed on the assumption that it is not a spurious coincidence,
but rather reflects the basic validity of the underlying physical model.

\section{Physical Properties of \lya Absorbers}\label{sec: phys}

The good agreement between the observed and simulated absorber statistics
suggests that the simulations are accurately describing the origin of
weak \lya absorbers at low redshift.  In this section, we investigate
the physical conditions in the simulated IGM that give rise to weak
\lya absorbers.  These studies are similar in spirit to those in
DHKW, but with a focus on the weak absorber population seen in
these STIS spectra.  We also investigate the relationship between the
$b$-parameter distribution and the temperature of the IGM.

\subsection{The Temperature-Density Relation}\label{sec: rhot}

At high redshift, there is a tight relationship between density $\rho$
and temperature $T$ in the gas producing the \lya forest, arising from
the balance between photoionization heating and adiabatic cooling due
to Hubble expansion \citep[e.g.][]{hg97}.  At low redshift, shocks have
heated a significant fraction of even diffuse intergalactic baryons above
the equilibrium photoionization temperatures, diluting this tight relation
\citep[e.g.][DHKW]{cen99}.  However, this shock-heated gas tends to have
a low ionization fraction, and hence is less likely to give rise to \lya
absorbers, while a substantial amount of baryonic matter remains in the
cool, diffuse IGM (see Figure~1 in D01).  Thus \lya absorbing gas may
still have a reasonably tight $\rho-T$ relation.

\begin{figure}
\epsscale{1.2}
\plotone{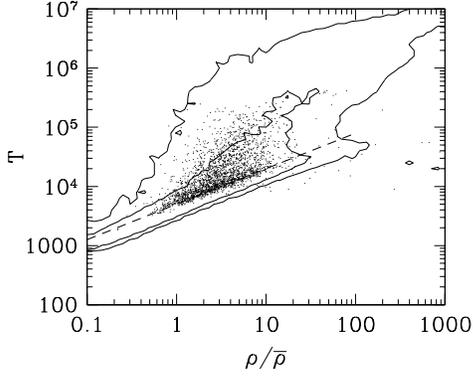}
\caption{Temperature vs. density for absorbers from our artificial
spectra.  The contours enclose 50\% and 90\% of all gas particles in our
simulation. These contours indicate that while an important shock-heated
population is present, the \lya absorbers preferentially sample gas in
the photoionized regime.  The dashed line shows a fit to the photoionized
absorbers, given in equation~(\ref{eqn: rhot}).  }
\end{figure}

Figure~4 shows the temperature $T$ vs. density $\rho/\bar\rho$ of all
absorbers identified in our 40 artificial spectra.  The density and
temperature of an absorber are taken to be those at the location 
of the absorber's maximum optical depth, and the
density is shown in units of the mean baryonic density at the absorber
redshift.  A significant fraction of the absorbers lie along a
relatively tight relation in $\rho-T$ space arising from
photoionization, but there are also many shock-heated absorbers at
higher $T$.  The photoionized absorbers follow the fitting formula
\begin{equation}\label{eqn: rhot}
T \approx 5000 \left( {\rho\over\bar\rho}\right) ^{0.6} \; {\rm K},
\end{equation}
shown as the dashed line, whose slope is similar to that seen at
high redshift \citep{cro98}.  For comparison, we show contours
encompassing 50\% and 90\% of all gas in our simulation.  In the
density range of STIS \lya absorbers, the absorbers' range of temperatures is
significantly smaller than that of all baryons.  Evidently, \lya
absorbers preferentially (though not uniquely) trace out gas in an
equilibrium photoionized state.

\subsection{Column Density vs. Gas Density}\label{sec: rhocol}

The FGPA indicates a power-law relationship between the optical depth
and the underlying gas density.  At low redshift, this relationship
becomes contaminated by hot gas that lies off the $\rho-T$ relation as
seen in Figure~4, resulting in absorbers with more widely varying
ionization fractions, and hence more widely varying underlying
densities.

\begin{figure}
\epsscale{1.2}
\plotone{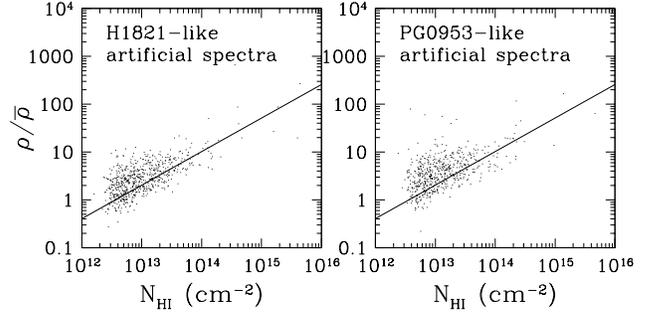}
\caption{Column density $\nh$ vs. density in units of the mean
$\rho/\bar\rho$, for absorbers in our artificial spectra.  Left and right
panels show results from H1821-like and PG0953-like artificial spectra,
respectively.  The line shows the relation given in equation~(\ref{eqn:
colspa}).  }
\end{figure}

Figure~5 show plots of column density $\nh$ vs. density of
underlying gas for the absorbers identified in the 40 artificial
spectra.  As suggested by Figure~10 of DHKW, there is
substantially greater scatter at low redshift as compared to high
redshift.  The line shows a similar fit to that in DHKW,
\begin{equation}\label{eqn: colspa}
{\rho \over \bar{\rho}}
\approx 12\; \left[ {\nh\over 10^{14}\cdunits}\right] ^{0.7}\; 10^{-0.4 z},
\end{equation}
corrected for the difference in $\ggh$ used (discussed further in
\S\ref{sec: ggh}).  This is a reasonable fit to the artificial STIS
absorbers, though there is a hint of greater scatter than in the $z=0$
absorber population of DHKW.  This may be because DHKW used S/N=30 per
pixel, whereas our STIS spectra have S/N$\sim 6-15$ per pixel, resulting
in larger errors in profile fitting to obtain $\nh$, especially at
lower column densities.  In Figure~5, the left and right panels show the
results from H1821-like and PG0953-like artificial spectra, respectively.
The most deviant outliers in density (at low column densities) arise
in the lower-S/N PG0953-like spectra, indicating that S/N issues may
be important.

Given the large scatter, it is difficult to unambigously assign an
underlying density to an absorber of a given $\nh$, though a trend
clearly exists.  While higher quality data may tighten this relation
somewhat, as of now it appears the FGPA is not nearly as tight a
relation at low redshift as compared to high-$z$.  Thus techniques such
as recovery of the mass power spectrum \citep{cro98} will face greater
challenges in this regime.

\subsection{The Temperature of the IGM}\label{sec: temp}

Since higher column density systems arise from higher density gas,
and higher density gas has higher temperatures, one would expect that
$b$ should be correlated with $\nh$.  In particular, the smallest $b$
parameters at a given $\nh$ should mainly reflect the temperature
of the underlying absorbers, as thermal broadening should dominate
in the narrowest lines.  This fact has been used at high redshift
to constrain the temperature and ``equation of state" \citep[i.e. the
$\rho-T$ relation,][]{hg97} of the IGM \citep{sch99,mcd00}, and a similar
argument has been extended to low redshift as well \citep{ric00}.  It is
conceivable that recovering the IGM temperature is actually easier at
low-$z$, because at high redshift, narrow lines can arise from Voigt
profile fitting the asymmetric wings of larger lines, and may be confused
for true narrow absorbers.   This occurs much less frequently in the
sparser low redshift forest.

\begin{figure}
\epsscale{1.2}
\plotone{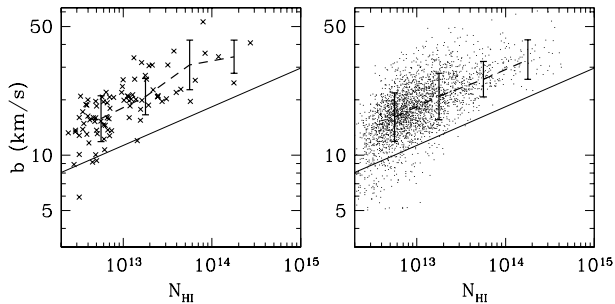}
\caption{$\nh$ vs. $b$-parameter for STIS-observed absorbers
(left panel) and absorbers in 40 artificial spectra (right panel).  The
thick dashed line shows the median $b$ in four bins of $\nh$, with variance
computed within each bin.  The solid line shows thermal broadening as a
function of $\nh$, given by equation~(\ref{eqn: bcol}).  }
\end{figure}

Figure~6 shows a plot of $\nh$ vs. $b$ for the absorbers identified in
PG0953+415 and H1821+643 (left panel) and absorbers in our artificial
spectra (right panel).  The dashed curve shows a running median of $b$
parameters with $\nh$, with variance in each bin.  There is a clear
trend for larger $b$-parameters at larger column densities, as would be
expected by the above argument.

We can quantify this relation as follows.  In \S\ref{sec: rhocol} we
confirmed that equation~(\ref{eqn: colspa}) is a reasonable fit to STIS
absorbers (though with large scatter).  Equation~(\ref{eqn:  rhot}) gives
our $\rho-T$ fit to photoionized absorbers, which are the coolest at any given
density.  Combining these relations and using $b=\sqrt{2k_BT/m_p}$ for
pure thermal broadening, we obtain an estimate for the minimum $b$
parameter as a function of column density and redshift:
\begin{equation}\label{eqn: bcol}
b_{\rm thermal} \approx 19\; \left[ {\nh\over 10^{14}\cdunits}\right] ^{0.21}\; 10^{-0.12z}\; 
{\rm km/s}.
\end{equation}
This relation is shown as the solid line in the two panels of Figure~6,
for $z=0.17$ (the mean redshift of the observed sample).  It is a
good approximation to the lower envelope of absorber $b$-parameters
for $\nh>10^{13.3}\cdunits$ in our artificial spectra (right panel).
Below this column density, presumably blending is more frequent, and
narrow lines from line wings dilute this relationship.  Also, noise can
cause some weak lines to appear narrower than they really are, since
weaker lines have larger profile fitting uncertainties.  As seen in
the left panel, equation~(\ref{eqn: bcol}) is consistent with a lower
envelope for all but the weakest STIS absorbers as well, though a larger sample
of absorbers will be required before such a relation may be recovered
directly from the observations.

Comparing the solid line (pure thermal broadening) with the dashed
line (median $b$) in Figure~6  suggests that thermal broadening is
responsible for a substantial portion of the absorber widths, with
$b_{\rm thermal}\approx 0.7 b$ for a typical absorber.  This fraction
remains fairly constant with $\nh$.  This is also seen in Figure~3,
where the dotted line shows the histogram of thermal widths for
$\nh>10^{13}\cdunits$ absorbers; its median value is $\sim 15$~km/s.
The contribution from thermal broadening is therefore greater than at
high redshift, where bulk flow broadening dominates \citep[e.g.][]{her96,
sch99}.

Equation~(\ref{eqn: bcol}) was obtained from fits to the particular
simulation used here; it is not true in general.  In order to measure
the temperature of the IGM unambiguously, it is important to test
equations~(\ref{eqn: rhot}) and (\ref{eqn: colspa}) over a wider
range of parameter space.  Nevertheless, the good agreement with the
$b$-parameter distribution, as well as the significant contribution
of thermal broadening to this distribution, suggests that these
simulations are accurately characterizing the ``equation of state"
of the low-$z$ IGM.  Our measurement is given by equation~(\ref{eqn:
rhot}), indicating a temperature of $\sim 5000$~K at the mean density.
Increasing the temperature by more than a factor of two would produce
thermal $b$-parameters alone that are larger than observed, so we
place an upper limit at $T(\bar\rho)\la 10^{4}$K.  Since this fit is
only for our current simulation, it is subject to the uncertainties
in modeling, but at least at high redshift measurements of $T$ are not
very sensitive to such uncertainties \citep{the00}.  Note that we are
measuring here the temperature of equilibrium photoionized \lya forest
absorbers only, and this does not exclude the possibility that the IGM
contains a component of hotter gas that may give rise to, say, \ion{O}{6}
absorption \citep{tri00b}.

A similar technique has already been used to determine the temperature of
the IGM from GHRS data, obtaining a value of $T(\bar\rho)=4700$~K for
$J_0=10^{-23}\junits$ \citep{ric00}, in agreement with our result.  With more
STIS data and better simulations, this type of analysis should be able
to robustly measure the temperature of the low-redshift IGM in the
near future.

\section{Preliminary Constraints on $\ggh$}\label{sec: ggh}

As stated in \S\ref{sec: stats}, the agreement between the observed and
simulated column density distribution amplitude is in principle a test of
the ionization state of the absorbing gas.  Because the absorbing gas for
$\nh\la 10^{14}\cdunits$ is optically thin, $\tau_{\rm Ly\alpha}\propto
\nh\propto \ggh^{-1}$, so changing the ionization fraction of \lya forest
absorbers will shift the simulated $f(\nh)$ in the horizontal direction,
and alter the agreement with observations.

It is straightforward to determine the shift in $\nh$ that produces the best agreement
with observations, then translate that into a value for $\ggh$.
This yields $\ggh = 10^{-13.3\pm 0.12}\; {\rm s}^{-1}$ at an average
redshift of $\bar{z}=0.17$, up by $\approx 0.2$~dex from the value
of $\ggh$ originally used to construct the artificial spectra.
Figure~7 shows this value of $\ggh$ (filled circle) in
relation to other predicted and observed values.  The error bars on
this value are discussed below.  The value derived here is close to
the ``matched LCDM" curve from DHKW (solid line), since it was this
$J_\nu$ that produced the good agreement with the observed $f(\nh)$
seen in Figure~2.  This value is even closer to that predicted from the
evolution of the quasar population by \citet[dashed line]{haa96}
and three-fourths of a similar prediction by \citet[dotted line]{far98} 
at $z=0.17$.  A recent calculation accounting for emission from QSOs,
Seyferts and starbursts \citep{shu99} is shown as the diamond at $z=0$,
and is quite consistent with our measurement.

\begin{figure}
\epsscale{1.2}
\plotone{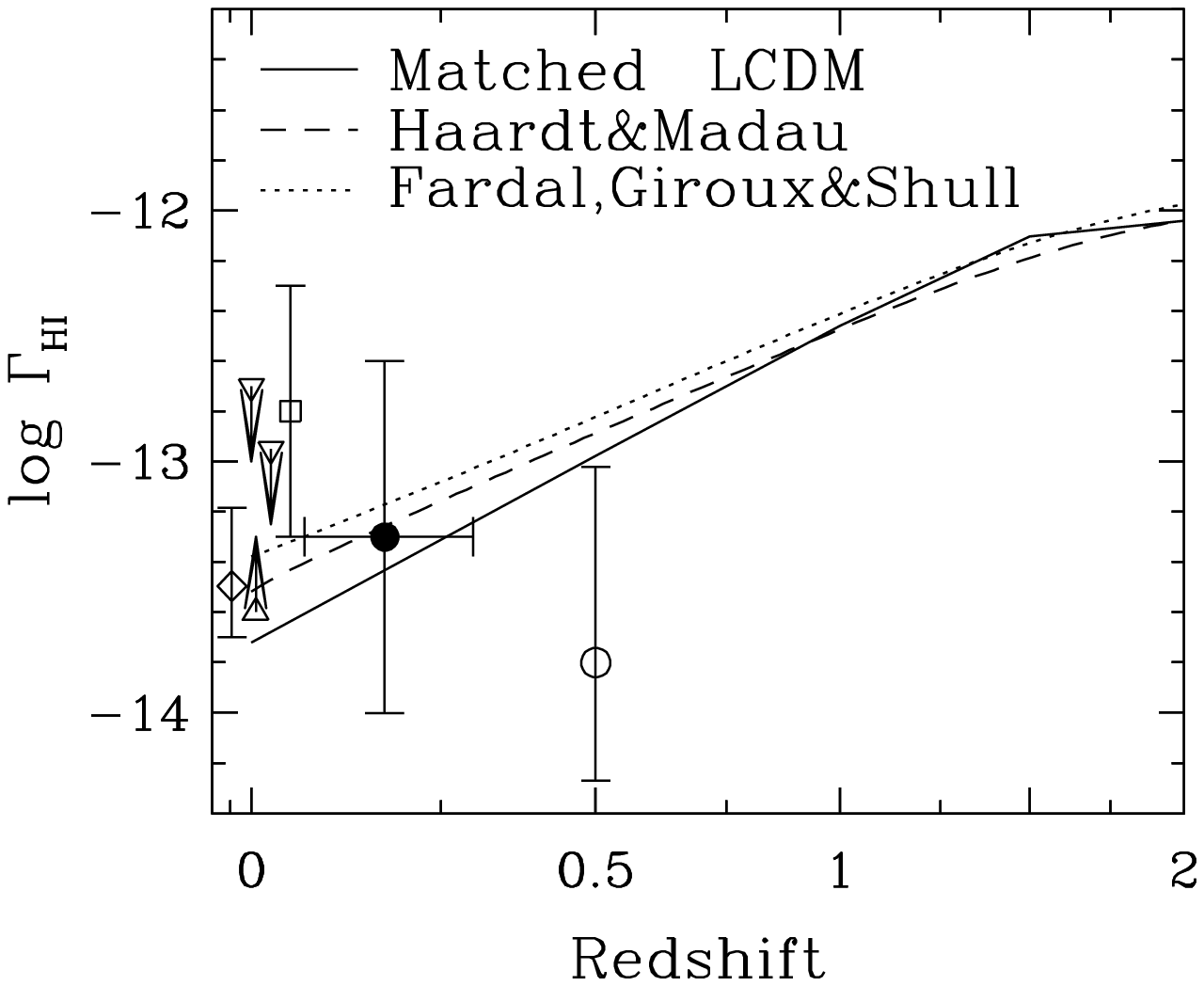}
\caption{Our measurement of $\ggh$ (filled circle) from the column density
distribution of the \lya forest.  The errors are dominated by modeling
uncertainties; the horizontal errors show the redshift range of absorbers
in our STIS spectra.  Predictions for the evolution of $\ggh$ are shown
as dashed \citep{haa96} and dotted \citep{far98} lines.  Solid line
shows the ``matched LCDM" rate that reproduces the evolution of $dN/dz$
for $W_r>0.24\AA$ FOS absorbers (DHKW).
This $\ggh$ was the one used to generate our artificial STIS spectra.
Various measurements are indicated: 
proximity effect \citep[open circle,][]{kul93};  
\ion{H}{1} truncation in spiral galaxy \citep[open square,][]{mal93,dov94};
photoevaporation off \ion{H}{1} cloud \citep[downward triangles, upper limits,][]{vog95,don95}; 
modeling of metal line ratios \citep[upward triangle, lower limit,][]{tum99}.  
Some points near $z=0$ have been offset slightly for clarity.}
\end{figure}

Figure~7 also shows various measurements that constrain $\ggh$ at low
redshift \citep[a more complete discussion may be found in][]{shu99}.  The
open circle at $z\sim 0.5$ comes from the proximity effect \citep{kul93},
the open square comes from modeling the radial truncation of \ion{H}{1}
in the isolated spiral NGC~3198 \citep{mal93,dov94}, and
the two downward triangles show upper limits from H$\alpha$ emission
due to photoevaporation off nearby \ion{H}{1} clouds \citep{vog95,don95}.  
An actual measurement using this last technique
is in preparation (R. Weymann, private communication).  \citet{tum99}
used metal absorption line ratios in the halo gas of NGC~3067 to model
the required photoionizing background, and their lower limit is shown
as the upward triangle.  Note that all these methods assume that the
ionizing radiation is metagalactic and not local, despite measuring the
field close to an observable object.  Conversely, weak \lya absorbers are
typically far from galaxies \citep{sto95,tri98}.  Additionally,
our statistical errors are smaller than those of any of the above
measurements, but as we now discuss, our modeling uncertainties dominate.

The FGPA (eq.~\ref{eqn: fgpa}) indicates that $\tau$ depends on the
density of the absorbing gas in addition to $\ggh$.  The exponent of this
dependence is approximately 1.6 at high redshift.  In the previous section
we showed that low redshift absorbers are consistent with this relation,
though the spread is significantly greater.  What is less certain is
how the density of the absorbing gas varies as one changes $\Omega_b$,
cosmology, and numerical parameters in the simulations.

At high redshift, nearly all the baryons are in the \lya forest, so
the FGPA can be used to directly probe $\Omega_b$ \citep{rau97,wei97}.
At low redshift, the fraction of gas giving rise to \lya absorbers is
significantly smaller \citep[$\sim 20\%$;][]{pen00}, because many
of the baryons have moved into other phases \citep[DHKW,][D01]{cen99}.  Though
modeling uncertainties are large (as we discuss below), the cosmological
simulations generally predict that at the present epoch roughly 1/3 of the
baryons are in diffuse photoionized gas, 1/3 are in the shock-heated IGM,
and 1/3 are in virialized systems such as galaxies and clusters (see D01
and references therein).  Studies of low-z \lya absorbers suggest that
the diffuse photoionized IGM contains a considerable fraction of the
baryons \citep{shu96,pen00}.  The shock-heated IGM is more challenging
to detect, but detections of \ion{O}{6} at low redshift suggest that the
predicted hot gas is indeed present \citep[e.g.][]{tri00b}, though since
\ion{O}{6} absorption can occur in photoionized gas and \lya lines can
originate in hot gas \citep[see, e.g., \S 4 in][]{tri00a}, there may be
some double-counting in estimates of baryon reservoirs traced by various
QSO absorbers.

Exactly how much baryonic matter remains in gas traced by \lya
absorbers is uncertain.  One estimate may be obtained if we consider
that low-$z$ absorbers preferentially arise in the so-called ``diffuse"
(i.e. photoionized) phase of DHKW and D01, as suggested by Figure~4.
In D01, the baryon fraction in this phase varies by as much as a factor of two
($\sim 20-40$\%) among their simulations having various volumes, resolutions,
and hydrodynamic algorithms.  Furthermore, those simulations all were
based on $\Lambda$CDM universes with similar $\Omega_b$.  While this world
model has gained much support recently \citep{bah99}, DHKW's analysis
of several different cosmologies also indicates a possible variation
of a factor of $\sim 1.5$ in the baryon fraction in this phase due to
the differing rates of structure formation.  Additionally, there is some
uncertainty in $\Omega_b$ itself, and uncertainties in $H_0$ come into
play as well since the observed quantity is usually the combination
$\Omega_b h^2$.  Due to these variations, combined with $\tau\propto
\rho^{1.6}/\ggh$, our best estimate for a $1\sigma$ systematic uncertainty in
the determination of $\ggh$ is roughly a factor of $\sim 5$.  Clearly this
dominates over the statistical error quoted above.

Thus we will claim a preliminary measurement of $\ggh = 10^{-13.3\pm 0.7}
{\rm s}^{-1}$ from the \lya forest, at $\bar{z}=0.17$.  This value is shown as the filled
circle in Figure~7; the horizontal error bar indicates the redshift range
covered by our STIS spectra.  For the spectral shape proposed by Fardal \etal
(similar to that of Haardt \& Madau and approximately $\nu^{-1.8}$), this
translates to $J_\nu=1.9\times 10^{-23}\junits$ at 912\AA.  Our preliminary
result would clearly benefit from improved simulations to better
quantify the uncertainties involved.  Unfortunately, analytic or
semi-analytic methods such as that of \citet{hui97} are not applicable
to this problem, since the physics of the diffuse IGM at low-$z$
involves a complex dynamical interplay of non-equilibrium structures.
A proper quantification would involve performing a suite of large-scale
hydrodynamic simulations exploring the relevant parameters, something
that is beyond our current computational capability.  Still, despite the
large and poorly quantified uncertainties in the value derived here,
ours is among the best observational constraints on the metagalactic
\ion{H}{1} photoionization rate to date, and offers hope that a more
precise determination may be forthcoming soon.

\section{Conclusions}\label{sec: disc}

We have examined the intergalactic \lya absorber population in STIS
spectra of PG0953+415 and H1821+643, and compared it to that in
artificial spectra drawn from a cosmological hydrodynamic simulation
of a $\Lambda$CDM universe.  The good spectral resolution (7~km/s) and
relatively high S/N of these STIS spectra yield an unprecedented view
into the weak absorber population at low redshift.  We find:

\begin{itemize}

\item The column density distributions agree quite well in slope and
amplitude.  The measured slope is $\beta= 2.04\pm 0.23$, considerably
steeper than that seen at high redshift ($\beta\approx 1.5$), indicating
that strong absorption line systems have evolved away faster than
weak ones.

\item The $b$-parameter (linewidth) distributions agree quite well,
with an intrinsic median $b$-parameter of $\sim 21$~km/s.  This value
is lower than that seen at high redshift, suggesting that the IGM has
cooled and/or bulk flow broadening has lessened.  The median $b$-parameter
increases with $\nh$, indicating that denser IGM gas is hotter.  From a
comparison of $b$-parameters with the expected thermal linewidth, thermal
broadening has an increased contribution to linewidths at low-$z$,
with $b_{\rm thermal}\approx 0.7 b$.

\item \lya absorbers preferentially arise in gas having temperatures
close to that expected from photoionization, with $T\propto
\rho^{0.6}$, though there is substantial scatter to higher
temperatures due to absorbers arising in the shock-heated IGM.

\item The density-column density relation, a.k.a. the fluctuating
Gunn-Peterson approximation, is considerably less tight at low redshift
than at high-$z$.  A trend still exists, consistent with that seen at
high redshift, but shows greater scatter.

\item The minimum $b$-parameter as a function of column density faithfully
yields the temperature-density relation in the IGM.  The simulation
analyzed here suggests $T\sim 5000$~K, with an upper limit of $10^4$~K,
for \lya absorbing gas at the cosmic mean density.

\item The amplitude of the column density distribution may be
used to place constraints on $\ggh$, the metagalactic \ion{H}{1}
photoionization rate at low-$z$.  We find, at $\bar{z}\approx 0.17$,
$\ggh\sim 10^{-13.3\pm 0.7}\; {\rm s}^{-1}$, with systematic uncertainties
in modeling being the dominant source of error.

\end{itemize}

In summary, weak \lya absorbers in the low-redshift IGM seen with
STIS appear to be physically similar to high redshift \lya absorbers, arising in
non-equilibrium large-scale structures that are highly photoionized by
the metagalactic UV flux, as suggested by DHKW.  Uncertainties
from our current simulation limit our ability to extract all possible
quantitative information contained even in this small sample of STIS
spectra, so improved simulations will be required to confirm the results
presented here.  Still, these preliminary results indicate that we
are close to understanding the low redshift \lya forest at a level
similar to what has been achieved at high redshift.  This would be a
remarkable achievement for studies of the nearby intergalactic medium
using ultraviolet spectroscopy, and analogous to what has transpired
recently with high-$z$ forest studies, promises to open up a new realm
for observational tests of theories of cosmology, structure formation,
and the evolution of intergalactic baryons.

\acknowledgments
We thank David Weinberg and Ray Weymann for helpful discussions. This
research made use of data reduction programs developed by the STIS IDT,
and we thank the STIS Team for allowing us to use this software.

This work was supported in part by NASA ATP grant NAG5-7066. Support
for this work was also provided by NASA through grant number
GO-08165.01-97A and Hubble Fellowship grant number GO-07465.01-A from
the Space Telescope Science Institute, which is operated by AURA, Inc.,
under NASA contract NAS5-26555.

\end{document}